\documentclass[10pt,conference]{IEEEtran}
\usepackage{cite,graphicx,psfrag,amsmath,amssymb,url}

\def\definedas{\triangleq}

\def\P{{\mathbb P}}
\def\lg{{\log_2}}
\def\Reals{{\mathbb R}}
\def\Reg{\ }
\def\Rno{}
\def\TM{\ }
\def\TMno{}
\newcommand{\defn}[0]{\textit}

\begin{document}
\title{On Conditional Branches in Optimal Decision Trees}
\author{\authorblockN{Michael B. Baer}
\authorblockA{Electronics for Imaging\\
303 Velocity Way\\
Foster City, California  94404  USA\\
Email: Michael.Baer@efi.com}}

\maketitle

\begin{abstract} 
The decision tree is one of the most fundamental programming
abstractions.  A commonly used type of decision tree is the alphabetic
binary tree, which uses (without loss of generality) ``less than''
versus ''greater than or equal to'' tests in order to determine one of
$n$ outcome events.  The process of finding an optimal alphabetic
binary tree for a known probability distribution on outcome events
usually has the underlying assumption that the cost (time) per
decision is uniform and thus independent of the outcome of the
decision.  This assumption, however, is incorrect in the case of
software to be optimized for a given microprocessor, e.g., in
compiling switch statements or in fine-tuning program bottlenecks.
The operation of the microprocessor generally means that the cost for
the more likely decision outcome can or will be less --- often far
less --- than the less likely decision outcome.  Here we formulate a
variety of $O(n^3)$-time $O(n^2)$-space dynamic programming algorithms
to solve such optimal binary decision tree problems, optimizing for
the behavior of processors with predictive branch capabilities, both
static and dynamic.  In the static case, we use existing results to
arrive at entropy-based performance bounds.  Solutions to this
formulation are often faster in practice than ``optimal'' decision
trees as formulated in the literature, and, for small problems, are
easily worth the extra complexity in finding the better solution.
This can be applied in fast implementation of decoding Huffman codes.
\end{abstract}

\section{Introduction}

Consider a problem of assigning grades to tests.  These tests might be
administered to humans or to objects, but in either case there are
grades $1$ through $n$ --- $n$ being $5$ in most academic systems ---
and the corresponding probabilities of each grade, $p(1)$ through
$p(n)$, can be assumed to be known; if unknown, they are assumed to be
identical.  Each grade is determined by taking the actual score, $a$,
dividing it by the maximum possible score, $b$, and seeing which of
$n$ distinct fixed intervals of the form $[v_{i-1}, v_i)$ the key
(ratio) $a/b$ lies in, where $v_0 = -\infty$ and $v_n = +\infty$.
This process is repeated for different values of $a$ and $b$ enough
times that it is worthwhile to consider the fastest manner in which to
determine these scores.

A straightforward manner of assigning scores would be to multiply (or
shift) $a$ by a constant $k$ ($\log_2 k$), divide this by $b$, and use
lookup tables on the scaled ratio.  However, division is a slow step
in most CPUs --- and not even a native operation in others --- and a
lookup table, if large, can take up valuable cache space.  The latter
problem can be solved by using a numerical comparisons to determine
the score, resulting in a \defn{binary decision tree} (also known as
an \defn{alphabetic binary tree}).  In fact, with this decision tree,
we can eliminate division altogether; instead of comparing scaled
ratio $ka/b$ with grade cutoff value, $v_i$, we can equivalently
compare $ka$ with $bv_i$, replacing the slow division of variable
integers with a fast multiplication of a variable and a fixed integer.
Depending on the application, this can be useful even if $b=1$ and no
division is required.  The only matter that remains is determining the
structure of the decision tree.

Such trees have a large variety of applications, including
nontechnical uses, such as the game of Twenty
Questions\cite[pp.~94--95]{CoTh} (also known as ``Yes and
No''\cite{Dic} or ``Bar-kochba''\cite{Reny}).  Technical uses includes
the compilation of switch (case) statements\cite{Sale, HeMe}.  An
optimized decision tree is known as an \defn{optimal alphabetic binary
tree}.

Often times these decision trees are hard coded into software for the
sake of efficiency, as in the high-speed low-memory \textsc{One-Shift}
Huffman decoding technique introduced in \cite{MoTu97} and illustrated
using C code in Fig.~2 of the same paper.  A shorter but similar
decision tree is illustrated in Fig.~\ref{pseudo} above by means of C
and assembly-like pseudocode.  We discuss this sample tree in
Section~\ref{static} of this paper, where a pictorial representation
of the tree is given as Fig.~\ref{branchtree}.

\begin{figure}
\centering
\begin{verbatim}
 if (V >= 34)      A. compare V, 34
                   B. branch to M if V<34
   if (V >= 42)    C. compare V, 42
                   D. branch to K if V<42
     if (V >= 65)  E. compare V, 65
                   F. branch to I if V<65
       P = 1;      G. P = 1
     else          H. go to N
       P = 2;      I. P = 2
   else            J. go to N
     P = 3;        K. P = 3
 else              L. go to N
   P = 4;          M. P = 4
                   N. end  
\end{verbatim}
\caption{Steps in a simple decision tree}
\label{pseudo}
\end{figure}

Algorithms used for finding such trees generally find trees with
minimum expected path length, or, equivalently, minimum expected
number of comparisons\cite{Knu71,HuTu,GaWa}.  We, however, want a tree
that results in minimum average run time, which is generally expressed
in terms of machine cycles, since these are usually constant time for a
given machine in a given mode.  The general assumption in finding an
optimal decision tree is that these goals are identical, that is, that
each decision (edge) takes the same amount of time (cost) as any
other; this is noted in Section 6.2.2 of Knuth's \textit{The Art of
Computer Programming}\cite[p.~429]{Knu3}.  In exercise 33 of Section
6.2.2, however, it is conceded that this is not strictly true; in the
first edition, the exercise asks for an algorithm for where there is
an inequity in cost between a fixed cost for a left branch and a fixed
cost for a right branch\cite{Knu31}, and, in the second edition, a
reference is given to such an algorithm\cite{Itai}.  Such an approach
has been extended to cases where each node has a possibly different,
but still fixed, asymmetry\cite{Shin}.

In practice the asymmetry of branches in a microprocessor is different
in character from any of the aforementioned formulations.  On complex
CPUs, such as those in the Pentium\Reg family, branches are predicted as
taken or untaken ahead of execution.  If the branch is predicted
correctly, operation continues smoothly and the branch itself takes
only the equivalent of one or two other instructions, as instructions
that would have been delayed by waiting for the branch outcome are
instead speculatively executed.  However, if the branch is improperly
predicted, a penalty for misprediction is incurred, as the results of
speculatively executed instructions must be discarded and the
processor returned to the state it was at prior to the branch, ready
to fetch the correct instruction stream\cite{HePa3}.  In the case of
the Pentium\Reg 4 processor, a mispredicted branch takes the
equivalent of dozens of instructions\cite{Fog}.  This penalty has only
increased with the deeper pipelines of more recent processors.  

In this paper, we discuss the construction of alphabetic binary trees
that are optimized with respect to the behavior of conditional
branches in microprocessors.  We introduce a general dynamic
programming approach, one applicable to such architecture families as
the Intel\Reg Pentium\Reg architectures, which use advanced dynamic
branch prediction, and the ARM\Reg architectures, most instances of which
use static branch prediction.
These are not only representative of two styles of branch prediction; they are
also by far the most popular processor architecture families for
32-bit personal computers and 32-bit embedded applications,
respectively.  ARM\Reg architectures such as those of the ARM7\TM and
ARM9\TM families use no or static branch prediction\cite{ArmPC}.  Such
processors are used for most mobile devices, including cell phones and
iPods\Rno.  (``ARM'' originally stood for ``Acorn RISC Machine,'' then
``Advanced RISC Machine,'' although now it is no longer considered an
acronym.)
Pentium\Reg designs and the XScale\TMno \cite{IntX} --- which is
viewed as the successor to ARM\Reg architecture StrongARM\Reg --- use
dynamic prediction.

Because the approach introduced here is more general than extant
alphabetical and search dynamic programming methods, using it to find
optimal decision trees is somewhat slower, having $O(n^3)$-time
$O(n^2)$-space performance.  This generality allows for different
costs (run times) for different comparisons due to such behaviors as
dynamic branch prediction and the use of conditional instructions
other than branches.  In the simplest case of static branch
prediction, entropy-based performance bounds are obtained based on
known results from related unequal edge-cost problems.  It should be
emphasized that the one-time $O(n^3)$-time $O(n^2)$-space cost of
optimization of these (usually small) problems is dwarfed by even the
slightest gain in repeated run-time performance.  The main
contribution is thus a method by which decision trees can be coded on
known hardware with minimum expected execution time.

\section{No prediction and static prediction}
\label{static}

It is easy to code the asymmetric bias of the branch for
implementations of static branch prediction.  In static prediction,
opcode or branch direction is used to determine whether or not a
branch is presumed taken, the most common rule being that forward
conditional branches are presumed taken and backward conditional
branches are presumed not taken\cite{HePa3}.  If the presumption is
satisfied, the branch takes a fixed number of cycles, while, if it is
not, it takes a greater fixed number of cycles.  Assume, for example,
that we want to use a forward branch, which is assumed not to be
taken.  We thus want the less likely outcome to be the costlier one,
that the branch is taken: If it is less likely than not
that the item is less than $v_i$, the branch instruction should
correspond to ``branch if less than $v_i$,'' as in all branches used
in Fig.~\ref{pseudo}.

\begin{figure}[t]
     \centering
     \psfrag{p1}{{\tiny $p(1)$}}
     \psfrag{p2}{{\tiny $p(2)$}}
     \psfrag{p3}{{\tiny $p(3)$}}
     \psfrag{p4}{{\tiny $p(4)$}}
     \psfrag{1}{$1$}
     \psfrag{3}{$3$}
     \psfrag{L}{{\tiny $b_1(1) = 0$}}
     \psfrag{R}{{\tiny $b_1(2) = b_1(3) = b_1(4) = 1$}}
     \psfrag{RL}{{\tiny $b_2(2) = 0$}}
     \psfrag{RR}{{\tiny $b_2(3) = b_2(4) = 1$}}
     \psfrag{RRL}{{\tiny $b_3(3) = 0$}}
     \psfrag{RRR}{{\tiny $b_3(4) = 1$}}
     \includegraphics{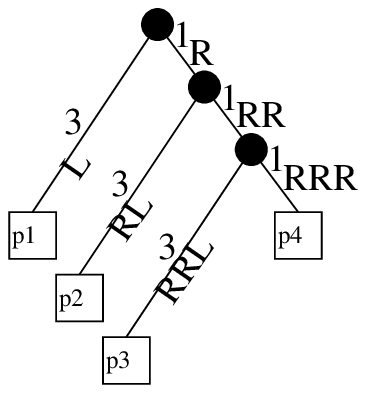}
     \caption{An optimal branch tree with edge costs for $c = (c_0~c_1) = (3~1)$}
     \label{branchtree}
\end{figure}

This \defn{branching problem}, applicable to problems with either no
true branch prediction or static branch prediction, considers positive
weights $c_0$ and $c_1$ such that the cost of a binary path with
predictability $b_1b_2 \cdots b_k$ is $\sum_{j=1}^k c_{b_j}$ where
$b_j = 0$ for a mispredicted result and $b_j = 1$ for a properly
predicted result.  Such tree paths are often pictorially illustrated
via longer edges on the corresponding tree, so that path depth
corresponds to path cost, e.g., Fig.~\ref{branchtree}.  This tree
corresponds to the C and pseudocode of Fig.~\ref{pseudo}.  The overall
expected cost (time) to minimize is
$$T_{p,c}(b) \definedas \sum_{i=1}^n p(i) \sum_{j=1}^{l(i)} c_{b_j(i)}$$
where $p(i)$ is the probability of the $i$th item, $l(i)$ is the
number of comparisons needed, and $b_j(i)$ is $0$ if the result of the
$j$th branch for item $i$ is contrary to the prediction and $1$ otherwise.

More formally, 

$$
\begin{array}{ll}
\mbox{Given } & p = (p(1)~p(2)~\ldots~p(n)),~p(i) > 0, \\
& \sum_i p(i) = 1 ;\\
& c_0,c_1 \in \Reals_+ \mbox{ such that } c_0 \geq c_1 . \\
\mbox{find } & B \mbox{, a full binary tree} ; \\
& b \mbox{, an assignment of costs to edges of} \\
& B \mbox{ such that each nonleaf is connected} \\
& \mbox{to its children by edges, one with cost} \\
& c_0 \mbox{, and the other with cost } c_1 . \\
\mbox{minimizing} & 
T_{p,c}(b) \definedas \sum_{i=1}^n p(i) \sum_{j=1}^{l(i)} c_{b_j(i)} \\
\mbox{where } & \mbox{the } j \mbox{th edge along the path from root} \\
& \mbox{to } i \mbox{th leaf is assigned cost } c_{b_j(i)} ; \\
&\mbox{the number of edges on the path from} \\
&\mbox{root to } i \mbox{th leaf is } l(i).
\end{array} 
$$

\begin{figure}[ht]
     \centering
     \psfrag{3}{{\small $3$}}
     \psfrag{5}{{\small $5$}}
     \resizebox{7cm}{!}{\includegraphics{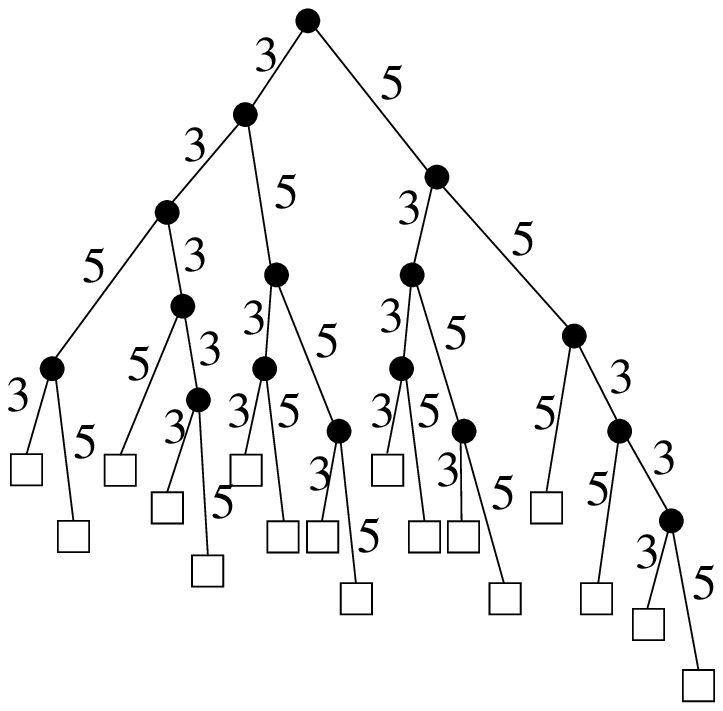}}
     \caption{Optimal branch tree for codeword lengths in optimal prefix coding of Zipf's law}
     \label{Zipfig}
\end{figure}

Sample representations are shown in Fig.~\ref{branchtree} and
Fig.~\ref{Zipfig}, the former being labeled with the values of
$b_j(i)$.  Again, to emphasize the total cost in this pictorial
representation, edges are portrayed with depth proportional to their
cost.  The cost (and thus depth) of leaf $3$ in
Fig.~\ref{branchtree} is, for example,
\begin{eqnarray*}
\sum_j c_{b_j(3)} &=& c_{b_1(3)}+c_{b_2(3)}+c_{b_3(3)} \\
&=& c_1 + c_1 + c_0 = 1 + 1 + 3 = 5.
\end{eqnarray*}  

Table~\ref{types} gives the context for this branching problem among
other binary tree optimization problems.  These other problems are
referred to as in the survey paper \cite{Abr01}.  In most problem
formulations, edge cost is fixed, and, where it is not fixed, edges
generally have costs according to their order, i.e., a left edge has
cost $c_0$ and a right edge has cost $c_1$.  Relaxing this edge-order
constraint in the unequal-cost alphabetic problem results in the
branching problem we are now considering.  Relaxing the alphabetic
constraint from either the original alphabetic problem or the
branching problem leads to Karp's nonalphabetic problem; since output
items in Karp's problem need not be in a given (e.g., alphabetical)
order, the tree optimal for the ordered-edge nonalphabetic problem is
also optimal for the unordered-edge nonalphabetic problem.

Thus the cost $T^{\mbox{\scriptsize Karp}}$ for the optimal tree under
Karp's formulation --- also called the \defn{lopsided tree problem}
--- is a lower bound on the cost of the optimal branch tree, whereas
the cost $T^{\mbox{\scriptsize Itai}}$ for the optimal tree under
Itai's (alphabetic) formulation is an upper bound on the cost of the
optimal branch tree.  This enables the use of bounds in \cite{AlMe}
--- including the lower bound originally formulated in \cite{Krau} ---
for the branching problem.  Specifically, if $b^{\mbox{\scriptsize
opt}}$ is the optimal branching function and $T^{\mbox{\scriptsize
opt}} = T_{p,c}(b^{\mbox{\scriptsize opt}})$ the associated cost for
the optimal tree, then
$$\frac{H(p)}{d} \stackrel{{\mbox{\tiny \cite{Krau}}}}{{\mbox{\tiny
$\leq$}}}{} T^{\mbox{\scriptsize Karp}} \leq T^{\mbox{\scriptsize
opt}} \leq T^{\mbox{\scriptsize Itai}} \stackrel{{\mbox{\tiny
\cite{AlMe}}}}{{\mbox{\tiny $\leq$}}}{} \frac{H(p)+1}{d} +
\max\left\{c_0,c_1\right\}$$ where $H$ is the entropy function $H(p) =
- \sum_i p(i) \lg p(i)$ and $d$ satisfies $2^{-dc_0}+2^{-dc_1}=1$.
If $\rho = c_0/c_1$ and $x$ is the sole
positive root of $x^{\rho}+x-1=0$, then $d = -c_1^{-1} \lg x$. Thus,
for example, when $c = (3~1)$, 
$$x = \sqrt[3]{\frac{1}{2}+\sqrt{\frac{31}{108}}} -
\sqrt[3]{-\frac{1}{2}+\sqrt{\frac{31}{108}}}$$ so $d = \lg x^{-1}
\approx 0.5515$ and $$T^{\mbox{\scriptsize opt}} \in [(1.813\ldots) H(p),
(1.813\ldots) H(p) + 4.813\ldots].$$  When $c = (2~1)$, $x = 1/\phi$ so
$d = \lg \phi$, where $\phi$ is the golden ratio, $\phi =
(\sqrt{5}+1)/2$.  These bounds can be used to estimate optimal
performance and determine whether or not to use a decision tree when
it is one of multiple implementation choices.

\begin{table}
\centering
\begin{tabular}{l||c|c|c}
&\multicolumn{3}{c}{\textit{edge cost/order restriction}} \\[6pt]
\textit{alphabetic?} &\textbf{Constant cost} &\textbf{Ordered} &\textbf{Unrestricted} \\[6pt]
\hline
\hline
&&& \\[-4pt]
\textbf{Yes}    &Hu-Tucker
  & Itai & branching problem \\
&\cite{GiMo,HuTu,GaWa,Knu3} & \cite{Itai,Shin} & (static) \\[6pt]
\hline
&& \multicolumn{2}{c}{} \\[-4pt]
\textbf{No} &Huffman\cite{Huff}
  & \multicolumn{2}{c}{Karp\cite{Karp,GoRo,BGLR}} \\[8pt]
\end{tabular}
\caption{Types of decision tree problems}
\label{types}
\end{table}

The key to constructing an optimizing algorithm is to note that any optimal
branching tree must have all its subtrees optimal; otherwise one could
substitute an optimal subtree for a suboptimal subtree, resulting in a
strict improvement in the result.  The branching problem is thus, to
use the terminology of \cite{VaPe}, \defn{subtree optimal}.  Each tree
(and subtree) can be defined by its \defn{splitting points}.  A
splitting point $s$ for the root of the tree means that all items
(grades) after $s$ and including $s$ will be in the right subtree
while all items before $s$ will be in the left subtree, as per the
convention in \cite{GiMo, Knu71, Knu3}.  
Since there are $n-1$ possible splitting points for the root, if we
know all potential optimal subtrees for all possible ranges, the
splitting point can be found through sequential search of the possible
combinations.  The optimal tree is thus found through dynamic
programming, and this algorithm has $O(n^3)$ time complexity and
$O(n^2)$ space complexity, in a similar manner to \cite{GiMo}.

The dynamic programming algorithm is relatively straightforward.  Each
possible optimal subtree for items $i$ through $j$ has an associated
cost, $c(i,j)$ and an associated probability $p(i,j)$; at the end,
$p(1,n)=1$ and $c(1,n)$ is the expected cost (run time) of the optimal
tree.

The base case and recurrence relation we use are similar to those of
\cite{Itai}.  Given unequal branch costs $c_0$ and $c_1$ and
probability mass function $p(\cdot)$ for $1$ through~$n$,
\begin{equation}
\begin{array}{rcl}
c(i,i) &=& 0 \\ 
c'(i,j) &=& \min_{s \in (i,j]} \{c_0 p(i,s-1) + c_1 p(s,j) +{} \\
&& \quad c(i,s-1) + c(s,j)\} \\
c''(i,j) &=& \min_{s \in (i,j]} \{c_1 p(i,s-1) + c_0 p(s,j) +{} \\
&& \quad c(i,s-1) + c(s,j)\} \\
c(i,j) &=& \min\left\{c'(i,j),c''(i,j)\right\}
\end{array}
\label{opt}
\end{equation}
where $p(i,j) = \sum_{k=i}^j p(i)$ can be calculated on the fly along
with~$c(i,j)$.  The last minimization determines which branch
condition to use (e.g., ``assume taken''
vs. ``assume untaken''), while the minimizing value of $s$ is the
splitting point for that subtree.  The branch condition to use ---
i.e., the bias of the branch --- must be coded explicitly or
implicitly in the software derived from the tree.  

Knuth\cite{Knu71} and Itai\cite{Itai} begin with similar algorithms,
then reduce complexity by using the property that the splitting point
of an optimal tree for their problems must be between the splitting
points of the two (possible) optimal subtrees of size $n-1$.  Note
that \cite{Itai} claims that this property can be extended to
nonbinary decisions, a claim that was later disproved in \cite{GoWo}.
The branching problem considered here also lacks this property.
Consider $p = (0.3~0.2~0.2~0.3)$ and $c = (3~1)$, for which optimal
trees split either at $2$, as in Fig.~\ref{branchtree}, or at $4$, the
mirror image of this tree.  In contrast, the two largest subtress, as
illustrated in the figure and its mirror image, both have optimal
splitting points at~$3$.

The optimal tree of Fig.~\ref{branchtree} is identical to the optimal
tree returned by Itai's algorithm for order-restricted
edges\cite{Itai}.  Consider a larger example in which this is not so,
the binomial distribution $p = (1\ 6\ 15\ 20\ 15\ 6\ 1)/128$ with $c =
(11~2)$.  If edge order is restricted as in \cite{Itai}, the 
optimal tree has an
expected cost of $15.109375$.  If we relax the restriction,
as in the problem under consideration here
the optimal method, has an expected cost of $12.984375$, a
$14\%$ improvement.

A practical application of this problem, involving a decision tree, is
encountered in implementation of the \textsc{One-Shift} Huffman
decoding technique introduced in \cite{MoTu97}.  This implementation
of optimal prefix coding is fastest for applications with little
memory or small caches.  
Where the \textsc{One-Shift} technique is the preferred technique, we
can apply the methods of this section to optimize the method's
decision tree.  In the implementation illustrated in \cite{MoTu97},
the decision tree is used to determine codeword lengths based on
32-bit keys.  The suggested ``optimal search'' strategy involves a
hard-coded decision tree in which branches occur if ``greater than or
equal to'' each splitting point; in most static branch schemes, this
would result in ``less than'' taking fewer cycles than ``greater than
or equal to,'' but the tree used in \cite{MoTu97} was found
assuming fixed branch costs\cite{Turp}.  Here we show that we can improve upon
this.

Consider the optimal prefix code for random variable $X$ drawn from the Zipf distribution with $n =
2^{16}$, that is, $\P[X = i]=1/(i \sum_{j=1}^n j^{-1})$ which is approximately
equal to the distribution of the $n$ most common words in the English
language\cite[p.~89]{Zipf}.  Using Huffman coding, one can find that
this code has codeword lengths, $\ell(X)$, between $4$ to $20$, with
the number of codewords of each size and the probability that the
codeword will be a certain size given by Table~\ref{zipf}.

\begin{table}
\begin{center}
\begin{tabular}{r|rl|rcl}
length ($\ell$)&\multicolumn{2}{c}{\# of codewords}&\multicolumn{3}{c}{$p(i)$} \\
\hline
4&$1$&($2^0$)&$\P[\ell(X)=4]$&=&$0.08570759$  \\
5&$2$&($2^1$)&$\P[\ell(X)=5]$&=&$0.07142299$  \\
6&$4$&($2^2$)&$\P[\ell(X)=6]$&=&$0.06509695$  \\
7&$8$&($2^3$)&$\P[\ell(X)=7]$&=&$0.06216987$  \\
8&$16$&($2^4$)&$\P[\ell(X)=8]$&=&$0.06076807$  \\
9&$32$&($2^5$)&$\P[\ell(X)=9]$&=&$0.06008280$  \\
10&$64$&($2^6$)&$\P[\ell(X)=10]$&=&$0.05974408$  \\
11&$128$&($2^7$)&$\P[\ell(X)=11]$&=&$0.05957570$  \\
12&$256$&($2^8$)&$\P[\ell(X)=12]$&=&$0.05949175$  \\
13&$512$&($2^9$)&$\P[\ell(X)=13]$&=&$0.05944984$  \\
14&$1024$&($2^{10}$)&$\P[\ell(X)=14]$&=&$0.05942890$  \\
15&$2048$&($2^{11}$)&$\P[\ell(X)=15]$&=&$0.05941844$  \\
16&$4096$&($2^{12}$)&$\P[\ell(X)=16]$&=&$0.05941321$  \\
17&$8192$&($2^{13}$)&$\P[\ell(X)=17]$&=&$0.05941059$  \\
18&$16384$&($2^{14}$)&$\P[\ell(X)=18]$&=&$0.05940928$  \\
19&$32747$&($2^{15}-1$)&$\P[\ell(X)=19]$&=&$0.05940732$  \\
20&$2$&&$\P[\ell(X)=20]$&=&$0.00000262$  \\
\end{tabular}
\end{center}
\caption{Distribution of Huffman codeword lengths for Zipf's law}
\label{zipf}
\end{table}

Consider a decision tree to find codeword lengths with an architecture
in which comparisons that result in untaken branches take $3$ cycles
(for both compare and branch), while comparisons that result in taken
branches take $5$ cycles.  This asymmetry, similar to that of many
ARM\Reg architectures, is small, but taking advantage of it results in
an improved tree.  This optimal tree, shown in Fig.~\ref{Zipfig},
takes an average of $15.93$ cycles, while the ``optimal search'' takes
an average of $16.44$ cycles.  This 3.1\% improvement, although not as
large as that of the binomial example, is still significant due to the
impact of the decision tree on overall algorithm speed.

\section{More advanced models}

With dynamic branch prediction\cite{HePa3}, which in more advanced
forms includes branch correlation, branches are predicted based on the
results of prior instances of the same and different branch
instructions.
This results in complex processor behavior.
Often several predictors will be used for the same branch instruction
instance; the predictor in a given iteration will be based on the
history of that branch instruction instance and/or other branches.  In
the problem we are concerned with, however, this does not result in as
many complications as one might expect; the probability of a given
branch outcome conditional on the branches that precede it is
identical to the probability of the branch outcome overall.  In the
case of previous branch outcomes for the same search instance ---
i.e., those of ancestors in the tree --- any given outcome is
conditioned on the same events --- i.e., the events that lead to the
branch being considered.  In the case of branches for previous items,
if items are independent, so are these branches.  In the case of
branches outside of the algorithm, these can also be assumed to be
either fixed given or independent of the current branch.

Thus, as long as each branch predictor is assigned at most one of the
decision tree branches, prediction can be modeled as a random process.
This process will result in each predictor converging to a stationary
distribution, which can be analyzed and optimized for.  
Simple analysis of the stationary distribution of a branch prediction
Markov chain, e.g., \cite{HPS}, can yield the expected time for a
given branch direction as a function of the probability of the branch.

Additional performance factors might include an additional asymmetry
between taken and untaken branches, the performance of branch target
buffers 
\cite{HePa3}, and differences among different comparison types.  For
example, if a $(<,\geq)$ comparison with a certain value has a smaller
cost than a comparison with another value --- say a comparison with a
power of two times a variable is faster due to reduced calculation
time --- then this can also be taken into account.  Similarly,
conditional instructions, often preferable to conditional branches,
can often be used, but only to eliminate a branch to leaves in the
decision tree.  Thus branches deciding between only two items might be
accounted differently than other branches.

With such a variety of coding options, there could be multiple
possible costs for any particular decision.  A general cost function
taking all this into account represents as $C_k(p',p'',i,j,s)$
the cost of choosing the $k$th of $m$ splitting methods for the step
necessary to split a subtree for items $[i,j]$ at splitting point $s$,
with splitting outcome probabilities $p'$ and $p''$.  (The most common
value for $m$ is $2$, the two choices being to assume a taken branch
versus to assume an untaken branch.)  The corresponding generalization
of (\ref{opt}) is:
\begin{equation*}
\begin{array}{rcl}
\displaystyle
c(i,i) &=& 0 \\[4pt]
c_k(i,j) &=& \displaystyle
\min_{s \in (i,j]} \{C_k(p(i,s-1),p(s,j),i,j,s) +{}\\
&& \quad 
c(i,s-1) + c(s,j)\}  \quad \forall k \\[4pt]
c(i,j) &=& \displaystyle \min_{k \in [1,m]} \left\{c_k(i,j)\right\} .
\end{array}
\end{equation*}
Once again, this is a simple matter of dynamic programming, and,
assuming all $C_k$ are calculable in constant time, this can be done
in $O(mn^3)$ time and $O(n^2 + n \log m)$ space, the $\log
m$ term accounting for recalculation and storage of the type of cost
function (decision method) used for each branch.  An even more general
version of this could take into account properties of subtrees other
than those already mentioned, but we do not consider this here.

\ifx \cyr \undefined \let \cyr = \relax \fi

\end{document}